\documentclass{phbauth}
\usepackage{graphicx}

\begin{document}

\begin{frontmatter}

\title{Superfluid critical temperature in 3D Fermi gas with repulsion}

\author[address1]{Dmitry V. Efremov\thanksref{thank1}},
\author[address1]{Maxim S. Mar'enko},
\author[address2]{Mikhail A. Baranov}
\author[address1]{Maxim Yu. Kagan},

\address[address1]{P.L.Kapitza Institute for Physical Problems, Ul. Kosygina 2,
Moscow 117334, Russia}
\address[address2]{Russian Research Center "Kurchatov Institute",
Kurchatov Square, Moscow, 123182, Russia}

\thanks[thank1]{Corresponding author.
 E-mail: efremov@kapitza.ras.ru}

\begin{abstract}
 The critical temperature of a  superfluid phase
transition in a Fermi gas with repulsive interaction is
found.  The influence of a magnetic field on the transition
is analyzed. The estimates for the critical temperature for
a trapped gas of $^6$Li atoms and $^3$He--$^4$He mixtures
are presented.

\end{abstract}


\end{frontmatter}

\def\ga{{\ \lower-1.2pt\vbox{\hbox{\rlap{$
          >$}\lower5pt\vbox{\hbox{$\sim$}}}}\ }}
\def\la{{\ \lower-1.2pt\vbox{\hbox{\rlap{$
          <$}\lower5pt\vbox{\hbox{$\sim$}}}}\ }}
\def\he3he4{$^3$He ¢ $^4$He}
\def\eps{\varepsilon}
\def\ua{\uparrow}
\def\da{\downarrow}
\def\su{\uparrow}
\def\sd{\downarrow}
\def\heee{$^3$He}
\def\heeee{$^4$He}
\def\ds{\displaystyle}
\def\g1t{$\tilde{\Gamma}_1$}

One of the most important questions in connection with
nonconventional superconductivity is the origin of an
attractive interaction. In this paper we show that a
 nonideal Fermi gas with purely repulsive bare interaction
 is unstable towards Cooper pairing with orbital
 momentum $l =1 $. This instability
 exists due to Kohn-Luttinger mechanism based on many-body
 effects \cite{Kohn}.
\section{Theoretical model}

 We consider a  nonideal Fermi-gas
 described by the Hamiltonian:

$$ \hat{H} =  \sum_{\sigma=\su,\sd \, {\bf p}} \xi_p
\hat{a}^{\dagger}_{{\bf p} \sigma}\hat{a}_{{\bf p} \sigma}+
g \sum_{ {\bf p}_i} \hat{a}^{\dagger}_{{\bf
p}1\su}\hat{a}^{\dagger}_{{\bf p}2\sd} \hat{a}_{{\bf
p}3\sd}\hat{a}_{{\bf p}4\su}, $$
 where $\xi_p = p^2/2m -\mu$, $\mu$ is the chemical potential, $g$ the
constant of a bare point-like repulsive interparticle
interaction, and different spin components, $\sigma = \su,
\sd$, are assumed  to have equal masses $m$ and
concentrations $n$.

As it was shown  in \cite{Kagan88,Fay68}, an effective
interparticle interaction originated from both the bare one
and many-body effects, is attractive when two particles
have a nonzero relative angular momentum $l$. This
attractive interaction is maximal for $l=1$ and, therefore,
results in a p-wave triplet Cooper pairing with the
critical temperature:
$$
T_{c1} = \tilde{\eps} \exp \left\{ -1/\nu_F |V^{eff}_1|
\right\} = \tilde{\eps} \exp \left\{ -12.9/ \lambda^2
\right\},
$$
where $\nu_F = mp_F/2\pi^2$ is the density of states at the
Fermi energy ($p_F = (6\pi^2n)^{1/3}$  the Fermi momentum
), $V_1^{eff} = (2 \ln 2 -1)g^2 \nu_F /5  $  the p-wave
harmonic of the effective interaction in the Cooper channel
(see \cite{Kagan88} for details), $a$ the scattering length
($a=mg /4\pi$ in the Born approximation), $\lambda=2 ap_F
/\pi$ the  gas parameter ($\lambda \ll 1$ for the
considered case), and $\tilde{\eps}$ the cutoff parameter
of the order of the Fermi-energy, $\tilde{\eps} \sim \eps_F
= p_F^2/2m$.

To fix the parameter $\tilde{\eps}$ and  find the critical
temperature  one has to keep all contributions up to
$\lambda^4$ in the Bethe-Salpeter equation, that defines
the critical temperature (see \cite{Efremov99} for more
details). These contributions originate from the effective
interaction, retardation effects (momentum and frequency
dependence of the effective interaction) and
renormalization of  Green functions (Z-factor and
$m^{*}$).

The corresponding critical temperature, found numerically,
is
\begin{small}
\begin{displaymath}\label{Tc_all:expand2}
  T_{c1} \approx \frac{2}{\pi} e^{C}\eps_F \exp \left\{
    -\frac{12.9}{\lambda^2(1+4.3\lambda)}+\frac{13.4}{(1+4.3\lambda)^2}
    \right\},
\end{displaymath}  \end{small}
where $C=0.577$ is the Euler constant, and neglected terms
are of order  $\lambda$. (Note, that this formula
extrapolates the expression for $T_{c1}$ from $\lambda \ll
1$ to $\lambda<1$.)

Magnetic field dependence of the critical temperature
$T_{c1}$ can be analyzed in the same way. We present the
results for $T_{c1}$ as a function of polarization
 $\alpha =
(n_{\su}-n_{\sd})/(n_{\su}+n_{\sd})$ on Fig 1. It turns
out, that the nonmonotonic dependence of $T_{c1}$ on
$\alpha$ is a result of a competition between the increase
of the angular dependence  of the effective interaction
$V^{eff}$ and decrease of its amplitude. (The former
increases $|V^{eff}_1|$ and, hence, $T_{c1}$, while the
later decreases.)

\begin{figure}[t]

\centerline{ %
\includegraphics[width=0.9\linewidth]{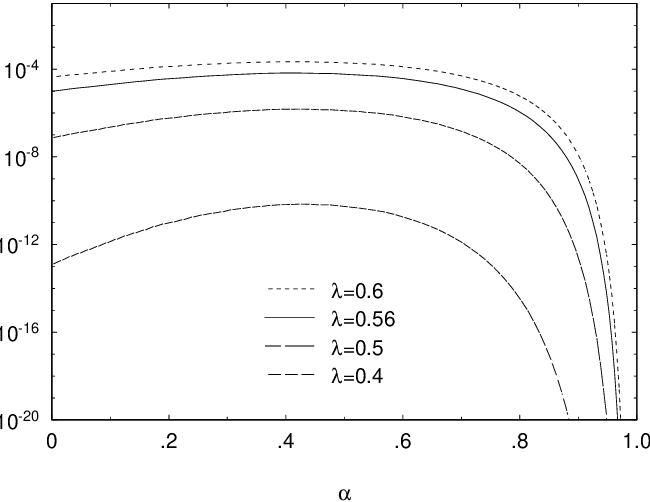}}
\label{fig:tclog0}
\caption{$T_{c1}/ \eps_F$ versus polarization
$\displaystyle \alpha $ for different $\lambda$. }
\end{figure}
\section{Conclusions.}
In conclusion,  let us mention two possible experimental
applications of the presented theory. The first one is to
 \heee-\heeee~ mixtures, providing that the concentration
$x$ of \heee~ is more than $3\%$. (In this case the
interaction between $^3$He atoms is repulsive.)
 The estimate  of the
critical temperature in zero magnetic field gives the value
$T_{c1} \approx 5\cdot10^{-6}$K for the maximal
concentration $x \approx 9.5\%$ (at pressure 10 bar). By
applying the magnetic field this value can be increased by
a factor of 6 (at polarization $\alpha \sim 40 \%$), that
gives a hope to observe the transition experimentally.

The second  application is to trapped neutral Fermi gases.
 For these systems the
critical temperature $T_{c1}$ is estimated to be of the
order of
 $10^{-7} \div 10^{-6}$K
for densities $n\sim 10^{14} \mbox{cm} ^{-3}$.
\begin{ack}

We acknowledge fruitful  discussions with A.F.~Andreev,
H.W.~Capel, I.A.~Fomin, Yu.~Kagan,  D.~Rainer, and
I.M~Suslov. This work was supported by  INTAS grant 98-963
and RFBR grants 98-02-17077 and 97-02-16532.
\end{ack}

\end{document}